# Field-free switching of a Spin-orbit torque device aided by interlayer-coupling induced domain walls

*Xiaotian Zhao, Wei Liu\*, Shangkun Li, Long Liu, Yuhang Song, Yang Li, Jun Ma, Xingdan Sun, Hanwen Wang, Xinguo Zhao and Zhidong Zhang*

Shenyang National Laboratory for Materials Science, Institute of Metal Research, Chinese Academy of Sciences, Shenyang 110016, China

E-mail: wliu@imr.ac.cn

Key Words: interlayer coupling, field-free, spin-orbit torque, synthetic antiferromagnetic, domain wall, spintronic

Abstract

The spin-orbit torque device is promising as a candidate for next generation magnetic memory, while the static in-plane field needed to induce deterministic switching is a main obstacle for its application in highly integrated circuits. Instead of introducing effective field into the device, in this work we present an alternative way to achieve the field-free current-driven magnetization switching. By adding Tb/Co multilayers at two ends of the current channel, assisting domain wall is created by interlayer exchange coupling. The field-free deterministic switching is achieved by the movement of the domain wall driven by current. By loop shift measurement we find the driven force exerted on the domain wall is determined by the direction of the in-plane moment in the domain wall. Finally, we modify the device into a synthetic antiferromagnetic structure to solve the problem of reading out signal. The present work broadens the choice of field-free spin-orbit torque device design and clearly depicts the difference between two different switching mechanism.

**Introduction**

Manipulating the magnetization by spin-orbit torque (SOT) in heavy metal/ferromagnet

(HM/FM) systems is a promising mechanism for magnetic random access memory (MRAM) application[1-3]. The 3-terminal SOT device is superior in the stability of the MgO tunnel barrier than the traditional 2-terminal spin-transfer torque (STT) device, and the switching efficiency can be further enhanced by increasing the spin-Hall angle[4] and interface transparency[5-7]. The application of SOT is not only important for data storage, but also lead to a new category of logic devices[8-10]. However, along with the SOT an external field parallel to the electric current is needed to achieve deterministic switching between two opposite magnetization directions, while keeping an external field in highly integrated circuits is impractical.

To exempt this field, different forms of effective field such as exchange bias[11-13], interlayer coupling[14, 15], gradient magnetic anisotropy[16, 17], Rashba field[18] or spin-transfer torque[19] etc have been attempted. In this exploration towards field-free SOT switching, the Dzyaloshinskii-Moriya interaction (DMI) plays a crucial role. DMI is a kind of indirect coupling between spins in ferromagnetic (FM) layer aided by spin-orbit-coupling induced by adjacent heavy metal layer in inversion-symmetry breaking structures. The DMI favors homochiral Neel-type domain wall (DW) in which the in-plane moment is aligned along the normal direction. According to the domain nucleation-expansion model of SOT switching[20], because of the chirality of the DW, both side of an enclosed domain under SOT have same moving direction, resulting in domain shift but not expansion. Different effective fields mentioned above are utilized to break the chirality, at least partially, to induce domain expansion which leads to deterministic switching.

Alternatively, if two ends of the current channel in the device can be initially magnetized in opposite directions, then the deterministic switching can be achieved by DW movement. In this approach, rather than be treated as a barrier needed to be overcome, the DMI is essential for stabilizing chiral Neel-type DW which can be effectively driven by SOT, similar to the racetrack devices[21-23].

Core of this scheme is how to keep the source areas of DW in fixed magnetization. Stray field induced by additional magnetic component[24, 25] and geometry effect[26] have been issued to solve this problem, but these solutions are either too complex for manufacture or hard to be realized when the device size decreases to tens of nanometer.

In this work we proposed an improved pinning scheme. A Ru/Pt bilayer is employed to connect Tb/Co pinning layers and Co/Ni/Co trilayer. Deterministic current-driven switching can be achieved once the magnetizations of Tb/Co pinning layers are set by field. The strong interlayer coupling makes it possible to investigate the DW assisted switching mechanism under the competition between DMI and external field. We found the direction of the in-plane spin in DW dominate the SOT-induced effective field. The geometry-independent nature and abundant expandability make our device design promising in practical manufacture and future spintronic research.

**Experimental Details**

The main fabrication procedures of the device are drawn in Figure 1(a). Stacks of Ta(20)/Pt(40)/Co(3)/Ni(6)/Co(3)/Ru(6.5)/Pt(3) (All thicknesses are given in Å) were deposited on thermally oxidized Si(001) substrate by DC magnetron sputtering with a base vacuum better than $4.0 \times 10^{-7}$ Torr. A relatively small silica layer thickness (50 nm) is chosen to avoid unwanted domain nucleation induced by the Joule heating of the current[27]. The less than 10 Å noble Ru/Pt capping layer is enough to prevent oxidization of the Co/Ni/Co FM layer. Meanwhile, the 6.5 Å thick Ru supplies antiferromagnetic interlayer coupling[28] and the 3 Å thick Pt can enhance the PMA of FM layer deposited in subsequent procedure[29]. During deposition, the Ar pressure was 2 mTorr and the deposition rates are all around 0.2 to 0.4 Å s$^{-1}$. Measured by The Co/Ni/Co tri-layer has saturated magnetization ($M_s$) of 750 emu cm$^{-3}$ and $H_\text{keff}$ = 8900 Oe. Then two sets of Tb/Co multilayer pinning layers were deposited in sequence by lift-off method. To form the pattern we employed a standard ultraviolet lithography technology. The first pinning layer(called Pad1) have the configuration of Co(8)/[Tb(10)/Co(7)]$_8$/ Au(100) and the second pinning layer(called Pad2) have the configuration of Co(8)/[Tb(7)/Co(7)]$_8$/Au(100). The thick gold layer is aimed at inducing large enough current shunting to keep the magnetization of pinned areas stable. Every time after the development of the photoresist, the sample was cleaned by oxygen-plasma for 20 s to ensure a good contact between Pt and Co for interlayer exchange coupling. Finally we patterned the stack into Hall-bar with 10 μm-wide current channel by Ar ion etching and deposited Ti(100)/Au(1000) electric pad on it, as shown in Figure 1(b).

The electric current is injected into the main channel of the device in two modes. In the measuring mode, sinusoidal current of 0.2 mA is applied to meaure the differential anormalous Hall resistance (AHE) $R_H$ of the device by Stanford SR830 Lock-in amplifier. To manipulate the magnetization of the device, the current mode was shifted to pulsing with duration of 1 ms. The source of both modes is an Agilent 32200 signal generator and a 10 MHz bandwidth bipolar amplifier. A home-made polar surface magneto optical Kerr (MOKE) microscope illuminated by green LED is employed to characterize the domain distribution. The loop shift method first introduced by Pai *et al*[30] is applied to evaluate the effective field generated by DMI on a vector magnet which can apply longitudinal and vertical magnetic field to the device at the same time. All the measurements and applied field directions follow the geometry noted in Figure 1(b).

**Results and discussion**

**Magnetic Properties**

To achieve desired lateral magnetic structure in the device, the $R_H$ values of the two areas pinned by Tb/Co multilayers versus $H_z$ are measured and shown in Figure 1(c). Both two sets of loops have sharp signal drops in the first and third quadrants, which is the typical feature of antiferromagnetic interlayer coupling. The larger coercivity for Pad1 comes from the less net magnetization compensated by thicker Tb layers, also indicating that the switching occurring at lower field for both loops should attribute to the Co/Ni/Co tri-layer. Minor loops of Pad1 and Pad2 have exchanging field $H_{ex}$ = 1904 Oe and 1758 Oe, corresponding to exchange strength $A_{Ru}$ = 1.71 and 1.58 mJ m$^{-2}$. The similar amplitude of of $A_{Ru}$ indicates that the oxygen-ion cleaning is effective to keep the good repeatability of our lift-off technology. Notably the small $R_H$ drops found in the central part of the $R_H$ -$H_z$ loops come from the switching of magnetization of the central cross of the anormal Hall bar, since the AHE signal is not so localized.

The different coercivity of the two pads give us a window to manipulate the remanent state of Co/Ni/Co in the two pinned areas by magnetization sequence. These different remanent states exert dramatic influence on the $R_H$ - $H_z$ loops of the central cross, as shown in Figure 1(d). The loops captured after saturated in +z (-z) direction has switching field in the saturated direction much larger than that of the opposite direction. Noting that the magnetization of the Co/Ni/Co tri-layer in remanent state is opposite to the saturation direction, this phenomenon implies that the pinned area supply switching assistance to the central area by exempting it from domain nucleation process, which has larger energy barrier than domain expansion. However, the bias can be ceased by adding an magnetizing field of ±2000 Oe opposite to the initial saturation direction. Diving into the coercivity window differs the magnetization directions of the two pinned areas and offers switching assistance to both magnetization directions of the central area. In the rest of this article, the remanent state after +6000 Oe and -2000 Oe magnetization is called State I and that after -6000 Oe and +2000 Oe magnetization is called State II for convenience.

**Field-free deterministic switch**

We cannot tell difference between State I and State II by hysteresis loops in Figure 1(d), but it is not the case for the switching process driven by SOT. Figure 2(a, b) shows the current-driven magnetization switching in the absence of external field. The $R_H$ values around $I_p = 0$ are the same as those shown in Figure 1d, making it credible that the magnetization of the central area is fully switched during the current-driven process. The critical current ($I_{cri}$) in both loops is 15.7 mA, corresponding to $1.93 \times 10^{11}$ A m$^{-2}$. Differential MOKE images are attached to the loops to present the domain structure during the current-driven switching sequence. In the case of State I shown in Figure 2(a), the reversal domain appear at the boundary of the Pad 2. During the expansion of this reversal domain, no nucleation of other domains is observed. After the sign of the current pulse come back to positive, the reversal domain start expansion from the opposite boundary, the same magnetization direction of the reversal domain and pinned area indicates that the pads are the source of the reversal domain. This suppose can be examined by changing the magnetization history to State II. As shown in Figure 2b, the magnetization can still be switched while the switching polarity is reversed. As shown in Figure 2(c, d), if there is only one cut-through DW in the current channel, the pinned areas guarantee an up-down ( ↑ ↓ ) DW under State I or down-up ( ↓ ↑ ) DW under State II. Considering that the Slonczewski-like SOT is directed along $\bm{m} \times (\bm{\sigma} \times \bm{m})$ [2, 23], where $\bm{m}$ is the moment and $\bm{\sigma}$ is the spin accumulation along $y$-axis generated by the SHE, the switching polarity is dependent on the $\bm{m}_{DW}$, defined as the in-plane $\bm{m}$ in the DWs, which show anti-clock chirality. This is consistent with previous works on Co/Ni/Co trilayer[22, 31]. Although the effective field of SOT is usually smaller than that of the PMA of the FM layer, the $\bm{m}_{DW}$ in the Neel-type DW can be switched to the $z$-axis since the torque generated by the PMA approaches 0 around the peak of the anisotropic potenial. Then the domain wall moving direction along with the reversal direction is determined by the nearer polar along z-axis to the in-plane spin in the Neel-type DW.

**Loop shift analysis**

To further understand the DW assisted switching it is necessary to investigate the effective field generated by the SOT. Compared to the harmonic method(Supplementary Note), the current assisted loop-shifting measurements[30, 32-34] is more appropriate in the existence of DWs. In this method, colinear static magnetic field $H_x$ and bias current $I_b$ are applied to the current channel of the Hall-bar during the measurement of $R_H$ values versus $H_z$. The expansion of the reversal domain will be assisted or suppressed by the SOT, depending on the relative directions of $I_b$, $H_x$ and $H_z$, resulting in loop shift. The slope of linear fitting derived from the relationship between the current densities and the loop shifts indicates the converting efficiency from electric current to SOT effective field under $H_x$. This method can be applied to our device, since the large coercivity of the Tb/Co ferrimagnet and the strong shunting effect of Au capping layer ensure that the magnetization configurations of the pinned areas can be stable against cooperation of current, in-plane field and moderate out-of-plane field.

For the DW assisted device in State I, loops measured with $H_x = 0$ in Figure 3(a) show non-zero shifts when $I_b \neq 0$, demonstrating the dependence of loop shift on the SOT effective field exerted on the assisting DW. Good linear dependence of loop shift on the $I_b$ implies no disturbance from the domain nucleation mechanism. Meanwhile, there is no obvious change of coercivity along with varying $I_{DC}$, implying the depinning of DWs is not so sensitive to Joule heating as the domain nucleation. By increasing $H_x$, the slope of the of loop-shifting curve efficiency monotonically decreases in the range of $H_x = 0$ Oe ~ 2000 Oe, presenting the competition between $H_x$ and $H_{DMI}$ on the alignment of $m_{DW}$. The opposition between $H_{DMI}$ and positive $H_x$ under State I is consistent with the results from Figure 2. The nonlinear curve for $H_x = 1400$ Oe is a sign of competition between two switching mechanism which is discussed in detail in Supplementary Note II. The curves of converting efficiency $\chi$ fitted from the loop shift curve versus $H_x$ over the range of ±3200 Oe for State I and II are shown in Figure 3(c). The intercept of the $\chi$-$H_x$ curves on x-axis give out the $H_{DM} = 1540$ Oe, i.e.

a DMI energy of 1.38 mJ m$^{-2}$. Around the maximum amplitude of $H_x$, two curves show the same saturated value of 9.36 Oe mA$^{-1}$. It suggests that no matter of the relative directions of $H_{DM}$ and $H_x$, $H_x$ of ±3200 Oe is large enough to overwhelm the alignment of $m_{DW}$. This value is less than value of 17.54 Oe mA$^{-1}$ obtained by the harmonic measurement (Supplementary Note), consistent with previous report[35].

Notably, although $H_x$ = 0 Oe and $H_x$ = 3080 Oe (-3080 Oe) have the same distance away from the compensated point of $H_{DMI}$ and $H_x$ under State I (State II), the absolute values of converting efficiency $|\chi|$ differ. $|\chi|$ gotten at zero field is 7.22 Oe mA$^{-1}$, about 77% of the $|\chi|$ values gotten at $H_x$ = 3080 Oe (-3080 Oe) for State I (II). To explain this phenomenon, we have to quantize the relationship between $\chi$ and $m_{DW}$ first. By rotating the sample holder, we make a contour map of the $|\chi|$ values among $H_x$- $H_y$ plane under State II, as shown in Figure 4(d). Around the margin of the contour map, where the combination of external field and DMI is large enough to align $m_{DW}$ in any direction, the $|\chi|/|\chi|_{max}$ value is highly coincident with $\cos|\alpha|$ value, $\alpha$ is the angle between the x-axis and the resultant vector of $H_{DM}$, $H_x$ and $H_y$. That means the projection of $m_{DW}$ along the current direction determine the ratio $|\chi|/|\chi|_{max}$. The only obvious deviation occurs at the margin sections around minimum $|\chi|$ values, implying that the large $H_y$ field possibly weaken the DMI by tilting the moment away from the normal direction of the domain. Supported by this conclusion, we suppose the insufficient $|\chi|$ value around zero field is because although the edges of the pinned areas are orthogonal to the x-axis, the DWs in the central area during switching process can hardly keep the shape of the edge for local pinning induced by structural inhomogeneity and nonuniform current distribution, just as the MOKE images shown in Figure 2(a, b). Since the $m_{DW}$ in Neel-type DW is normal to the tangent line, the sections of the DWs tilted from the original direction can only supply partial of the maximum driving force depending on the projection of $m_{DW}$ on the x-axis. If we assume the DW is only pinned at the boundary of the current channel[26, 36] and has a simplified semicircle profile, it is easy to deduce that the ratio between the driven forces of the curved profile and the original profile equals the area ratio between a circle and its circumscribed square, i.e. $\pi/4 \approx 78.5\%$, which is close to the experimental value. In contrast to DMI, the $H_x$ field align the moment along the x-axis regardless of the DW profile. That difference also contributes to the slope discontinuity of the $\chi$-$H_x$ curves around zero field.

We also fabricated reference Hall-bar without the Tb/Co pads on the same substrate and measured the $\chi$-$H_x$ curve, which is also attached in Figure 3(c). This curve shows saturating trend around 1500 Oe but don't have a sharp critical point. Referring to the intersections between the DW assisted $\chi$-$H_x$ curves and the x-axis, the point where the fitting line of the saturated parts intersect the slope line at origin point is a best approximation for $H_{DMI}$ definition. It is interesting to see that although the reference $\chi$-$H_x$ curve have the same saturated value with the other two curves, but its shape is different from the mid-value line of the two. This mid-value line is drawn as the green dash line in Figure 3(c). Indeed in the work of Mann *et. al.*[35] we can find a transition from the shape similar to this mid-value line to the more common shape of the data line of the reference device, along with decreasing $H_{DMI}$ and increasing $\chi$. Since an enclosed small domain with strong DMI can be seen as a topologically protected skyrmion[37], it implies when DMI is strong enough to suppress domain nucleation, the domain expansion may only start from both ends of the current channel, and the switching is dominated by the competition between ↑↓ and ↓↑ DWs, resulting in $\chi$-$H_x$ curve like the green dash line. The DW assisted device shows us a new perspective to analysis complex situation about SOT, which is specially helpful in the design of new spintronic devices.

In the device fabrication, the Ru/Pt composite layer is used as oxidation barrier layer after step 1 and exchange coupling interlayer in step 2 and 3, finally it further help us to build a synthetic

antiferromagnetic(SAF) structure in the central area to solve the problem of signal reading out in practical use of spintronics. In many device design for the field-free SOT-switching, two surfaces of the FM layer are both occupied to fulfil the designed function, leaving no space for tunnelling layer like MgO for tunnelling magnetic resistance(TMR) reading out. In our device, however, an additional FM layer can be capped on the central area to form the SAF structure and the top surface of the added FM layer is spared. To fabricate the SAF devices, we made another stack and added deposition of Co(4-13 Å)/Pt(10 Å) between step 3 and step 4 in Figure 1(a). The Pt capping layer is added to prevent oxidation. The $R_H$-$H_z$ loops measured from the device with $t_{Co}$ = 7.5 Å shows an antiferromagnetic curve with an additional switching between two SAF coupled states, as indicated in Figure 4(a) (Supplementary Note) [38]. The State I and II defined above is also used for this device, and the $R_H$-$I_p$ loops is shown in Figure 4(b). The difference between two $R_H$ levels corresponds to the SAF coupled states in Figure 4(a), indicating the competence of the DWs to lead the deterministic switching of the SAF structure, although along with critical current increased. The critical current shows monotonically positive correlation to the Co layer thickness, as shown in Figure 4(c). That indicates the DW movement in the device is in the regime of creep[39], and the mechanism of exchange torque used to explain the speed enhancement around compensation point for the SAF multilayer[31] does not suit the situation of this work. The feasibility of driving SAF magnetization by current allows us to design a magnetic tunnelling junction (MTJ) in which the capping Pt is replaced by oxide tunnelling layer and the reference magnetic layer is added at the top, as illustrated in Figure 4(d). In this design each magnetic component is either SAF or ferrimagnetic multilayers[40] hence in favour the reduction of stray fields.

As a conclusion, we have proposed a kind of field-free current-driven magnetization switching device with Tb/Co pads for creating the assisting DW. Evidenced by shift efficiency measurement and MOKE imaging, we found that the magnetization switching is dominated by assisting DWs occurred at the boundary of the pinned areas. The SOT efficiency is mainly determined by the orientation of the in-plane spins in the assisting DWs. The Ru/Pt surface can be used to build SAF structure in the central area for sparing a FM surface for reading out the TMR signal. This scheme is both applicable for the magnetization switching scheme and racetrack DW movement scheme, and the possibility of driving ferrimagnetic Tb/Co multilayers by all-optical helical switching[41, 42] boosts the device into a logic unit which can be reconfigured by femtosecond laser. Rather than a pure magnetic memory, the device design presented in this work is more like a multidiscipline spintronic platform.

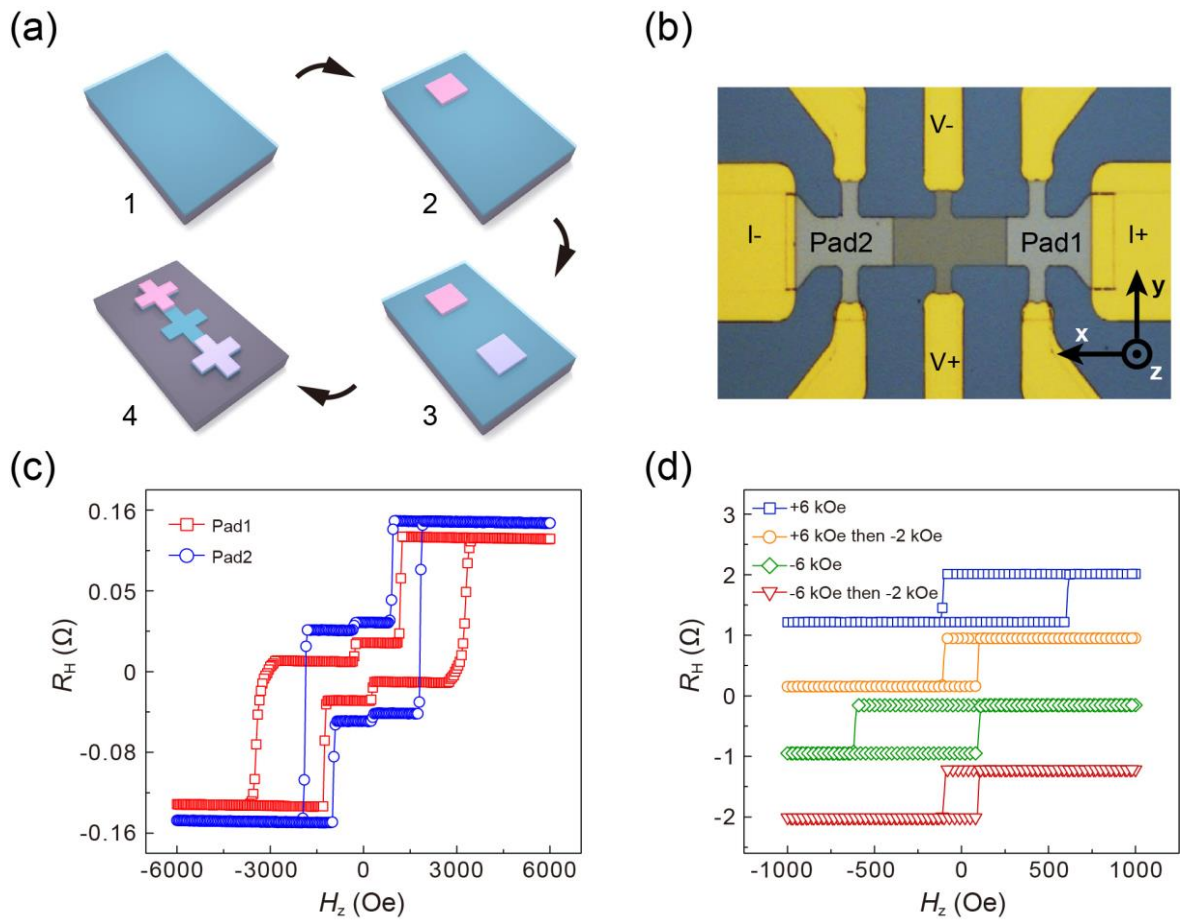

Figure 1. Structure and magnetic properties of the DW assisted device. (a) Main fabrication procedures of the DW assisted device. (b) The photography and transportation measuring configuration of the DW assisted device. (c) $R_H$ vs $H_z$ loops of Pad1 and Pad2. (d) $R_H$ vs $H_z$ loops of the central area measured after different magnetization histories.

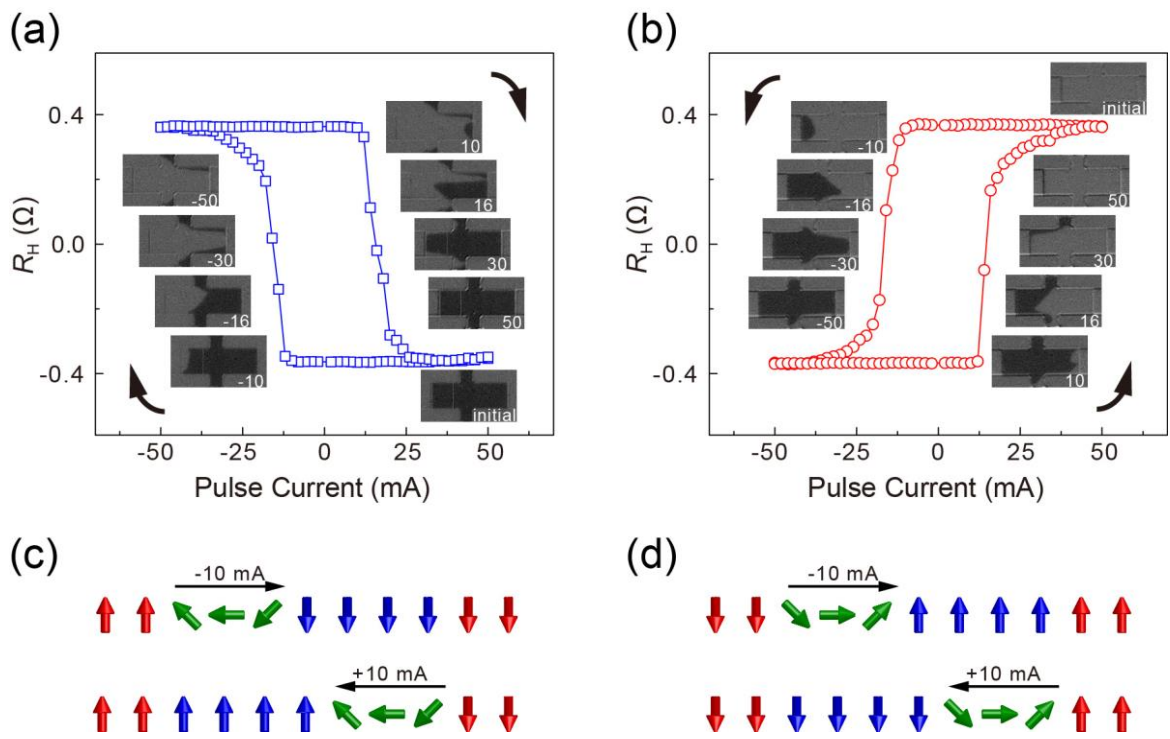

Figure 2. Field-free switching assisted by the DWs. (a, b) Field-free current-driven switching $R_H$ vs $I_p$ loop measured under State I (a) and State II (b) with MOKE images taken under labelled current densities. (c, d) The schematic illustrations of moment configuration of the Co/Ni/Co trilayer under State I (a) and State II (b) at the moments that the DWs tend to move along with the current direction labelled by black arrows, different colors represent moments pinned by the pads (red), uniformly magnetized in the central area (blue) and in the assisting DWs (green).

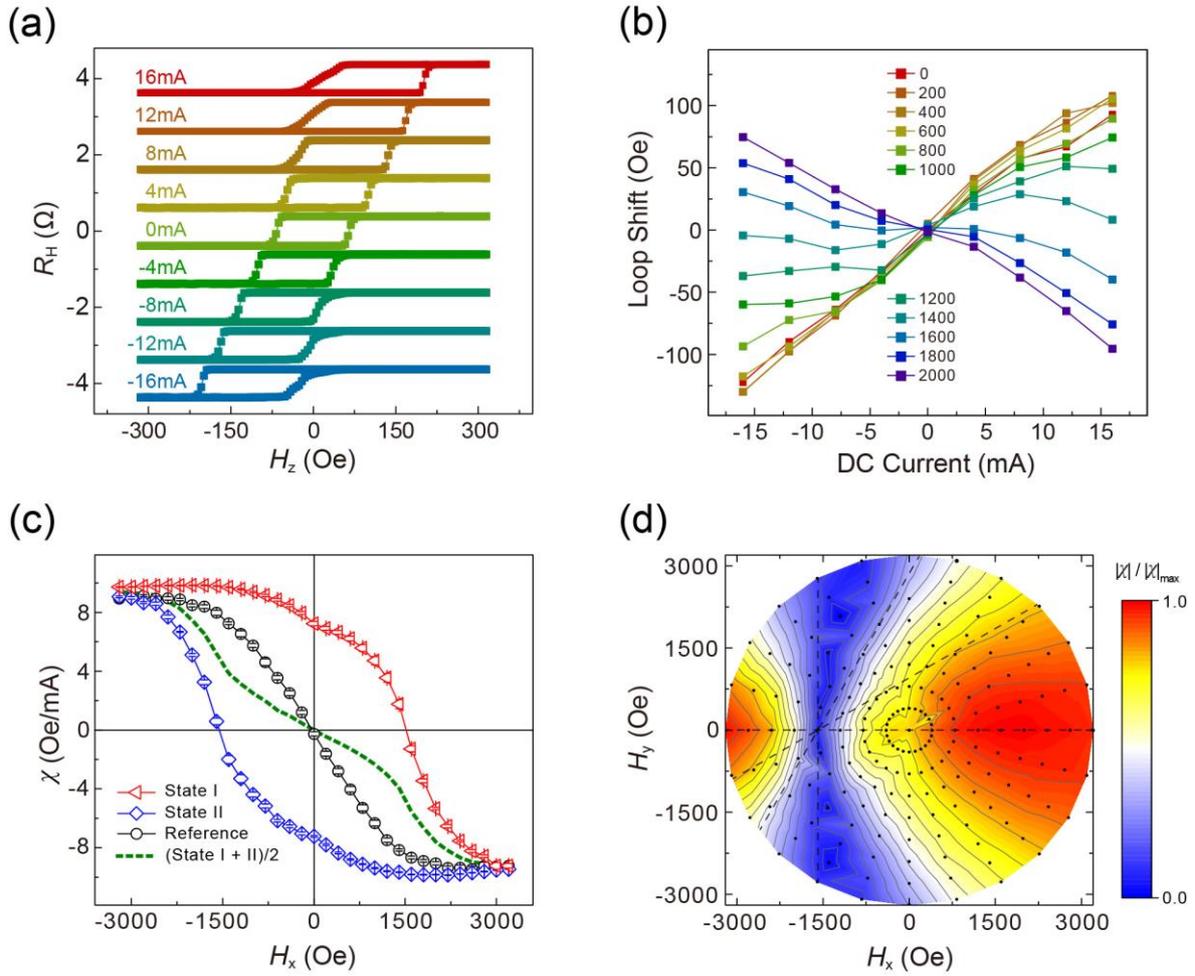

**Figure 3. SOT efficiency measured by the loop shifting method.** (a) Shifts of $R_H$ vs $H_z$ loops of the central area measured at zero field under State I along with different bias current. (b) Curves of Loop shift vs bias current measured under different $H_x$. (c) $\chi$-$H_x$ curves of the DW assisted device measured under State I (blue open diamond), State II (red open triangle), the average of the two curves (the green dash line), and $\chi$-$H_x$ curve of the reference device (black open circle). (d) Normalized $|\chi|$ contour map among $H_x$ and $H_y$ in State I, the dash lines are guild to the eyes for the boundary $|\chi|$ values on the same straight line which pass through the point (1530 Oe, 0 Oe) on $H_x$ - $H_y$ coordinate.

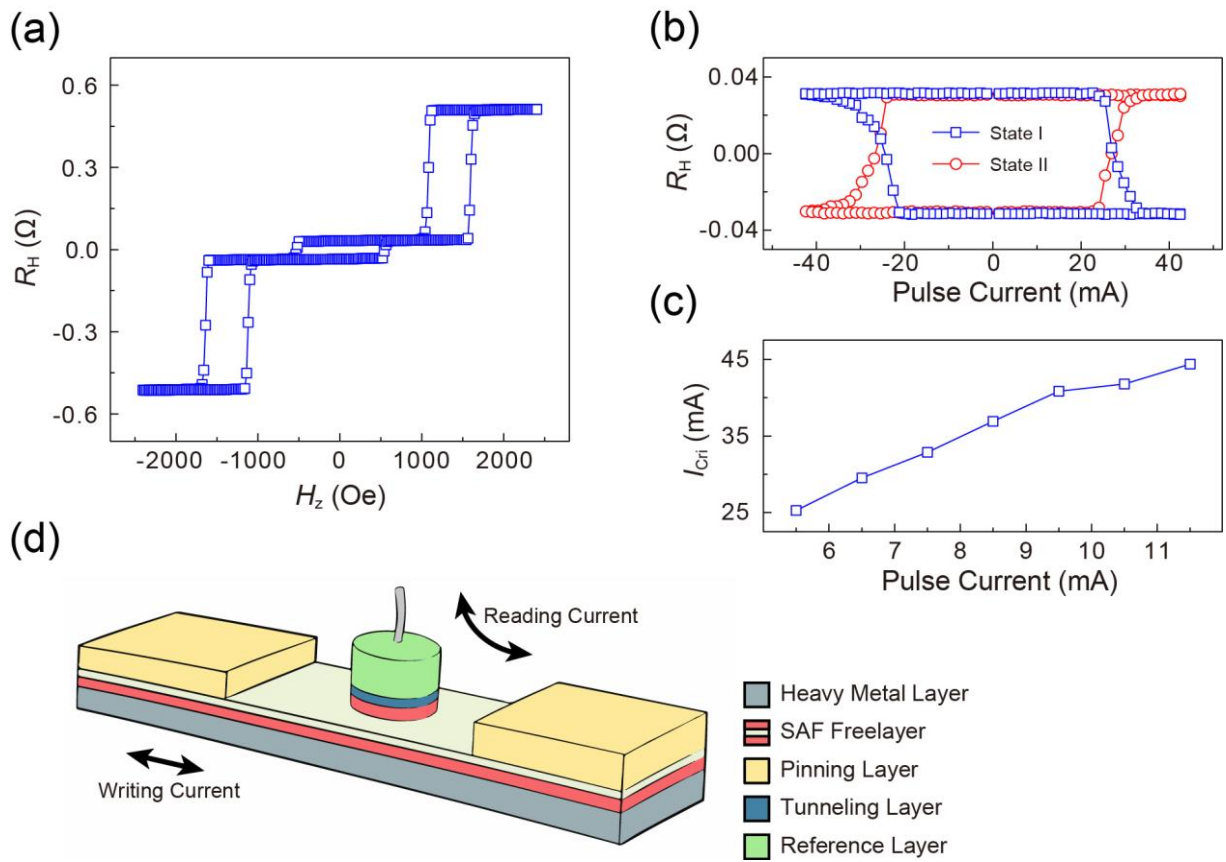

Figure 4 The SAF structure device design. (a) $R_H$ vs $H_z$ loop and (b) $R_H$ vs $I_p$ loops under State I and II of of the SAF device with $t_{Co}$ = 7.5 Å. (c) The $I_{cri}$ dependence on $t_{Co}$ for the SAF devices. (d) Schematic illustration for the MRAM design utilizing the DW assisted SAF structure.